\documentclass[aps,prl,twocolumn,showpacs, superscriptaddress, floatfix]{revtex4}
\usepackage{graphicx}
\usepackage{color}
\usepackage{float}
\usepackage{rotating}

\begin{document}
\title{Polaronic transport induced by competing interfacial magnetic order in a La$_{0.7}$Ca$_{0.3}$MnO$_{3}$/BiFeO$_{3}$ heterostructure}

\author{Y. M. Sheu}
\email{ymsheu@lanl.gov}
\affiliation{Center for Integrated Nanotechnologies, MS K771, Los Alamos National Laboratory, Los Alamos, New Mexico 87545, USA}
\author{S. A. Trugman}
\affiliation{Center for Integrated Nanotechnologies, MS K771, Los Alamos National Laboratory, Los Alamos, New Mexico 87545, USA}
\author{L. Yan}
\affiliation{Center for Integrated Nanotechnologies, MS K771, Los Alamos National Laboratory, Los Alamos, New Mexico 87545, USA}
\author{J. Qi}
\affiliation{The Peac Institute of Multiscale Sciences, Chengdu, Sichuan 610225, People's Republic of China}
\author{Q. X. Jia}
\affiliation{Center for Integrated Nanotechnologies, MS K771, Los Alamos National Laboratory, Los Alamos, New Mexico 87545, USA}
\author{A. J. Taylor}
\affiliation{Center for Integrated Nanotechnologies, MS K771, Los Alamos National Laboratory, Los Alamos, New Mexico 87545, USA}
\author{R. P. Prasankumar}
\email{rpprasan@lanl.gov}
\affiliation{Center for Integrated Nanotechnologies, MS K771, Los Alamos National Laboratory, Los Alamos, New Mexico 87545, USA}


\begin{abstract}
Using ultrafast optical spectroscopy, we show that polaronic behavior associated with interfacial antiferromagnetic order is likely the origin of tunable magnetotransport upon switching the ferroelectric polarity in a La$_{0.7}$Ca$_{0.3}$MnO$_{3}$/BiFeO$_{3}$ (LCMO/BFO) heterostructure. This is revealed through the difference in dynamic spectral weight transfer between LCMO and LCMO/BFO at low temperatures, which indicates that transport in LCMO/BFO is polaronic in nature.  This polaronic feature in LCMO/BFO decreases in relatively high magnetic fields due to the increased spin alignment, while no discernible change is found in the LCMO film at low temperatures. These results thus shed new light on the intrinsic mechanisms governing magnetoelectric coupling in this heterostructure, potentially offering a new route to enhancing multiferroic functionality.
\end{abstract}
 \maketitle

The quest to achieve strong magnetoelectric (ME) coupling has driven the surge in research on multiferroic materials over the past decade. However, this has been quite difficult to accomplish using bulk materials, motivating researchers to explore other approaches, most notably the use of transition metal oxide heterostructures \cite{Eerenstein2007NM,Wu2010NM,Yu2012MatToday}. In these novel systems, different degrees of freedom (DOFs) (e.g., charge, spin, and orbital) are coupled at a single interface between two different oxide layers to form a new state that displays properties dramatically different from those of the individual layers \cite{Heber2009Nature, Hwang2012NM, Ueda1998Science, Ohtomo2004Nature, Chakhalian2007Science}. Particular attention has been given to the coupling between ferromagnetic (FM), antiferromagnetic (AFM) and ferroelectric (FE) orders, as this could reveal new routes to realizing strong ME coupling \cite{Bea2006APL,Bea2006APLEB,Eerenstein2007NM,Hambe2010AM,Wu2010NM,Yu2012MatToday}.

Heterostructures consisting of manganite and multiferroic layers are particularly promising in this regard. The most extensively studied combination  \cite{Bea2006APL,Bea2006APLEB,Borisevich2010PRL,Wu2010NM,Yu2012MatToday} consists of the colossal magnetoresistive (CMR) manganite La$_{0.7}$Sr$_{0.3}$MnO$_{3}$ (LSMO) (or the similar compound La$_{0.7}$Ca$_{0.3}$MnO$_{3}$ (LCMO)), which is ferromagnetic below a critical temperature $T_c$, and the canonical multiferroic BiFeO$_{3}$ (BFO), which has coexisting coupled AFM and FE phases in which the magnetization can be switched by an applied electric (E) field \cite{Zhao2006NM}. The combination of these materials thus has great potential for exhibiting novel phenomena by coupling different FM, AFM, and FE phases across the interface. Indeed, a new interfacial state between LSMO and BFO has been discovered experimentally and discussed theoretically \cite{Wu2010NM,Yu2010PRL,Calderon2011PRB,Wu2013PRL}.  This state displays an exchange-bias (EB) field and magnetotransport that can be tuned by switching the FE polarization, increasing the potential for device control through interfacial coupling. However, a complete picture of the interplay between FM and AFM orders in this heterostructure has yet to be reached.

Current understanding of the EB field, which arises from the interaction between FM and G-type AFM (AFM(G))orders at the interface, is based on spin canting or pinning in BFO, with the assumption of negligible canting in the manganite layer \cite{Yu2010PRL,Calderon2011PRB,Wu2013PRL}.  However, given that the spin interaction is mutual, it is not clear why FM spins in LSMO can induce AFM spins in BFO to cant, but the reverse has not been discussed. In fact, the interfacial spin-spin interaction $J_{\text{FM-AFM(G)}}$ could enable the use of  effective 'AFM staggered fields' to control FM spins in an alternating arrangement, which cannot be done using an applied magnetic (B) field. Furthermore, relatively little effort has been made to explain the E-field switchable magnetotransport in the manganite layer, which is arguably of equal importance.

One can shed light on these issues using ultrafast optical spectroscopy (UOS), which has been demonstrated to be a sensitive probe of the charge, spin and orbital order in CMR manganites \cite {Averitt2001PRL, Lobad2001PRB, Tobey2008PRL,  Okimoto2007JPSJ, Prasankumar2007PRB,Ren2008PRB,Li2013Nature}. In particular, much insight into the physics of these systems has been obtained from probing large photoinduced changes in the optical conductivity between low and high frequencies (known as 'dynamical spectral weight transfer' (DSWT)) \cite{Averitt2001PRL,Lobad2001PRB,Ogasawar2003PRB,Tobey2008PRL,Okimoto2007JPSJ}, which are strongly coupled to the magnetotransport properties \cite{Millis1996PRB}.

\begin{figure*}[tb]
\begin{center}
\includegraphics[width=6.8in]{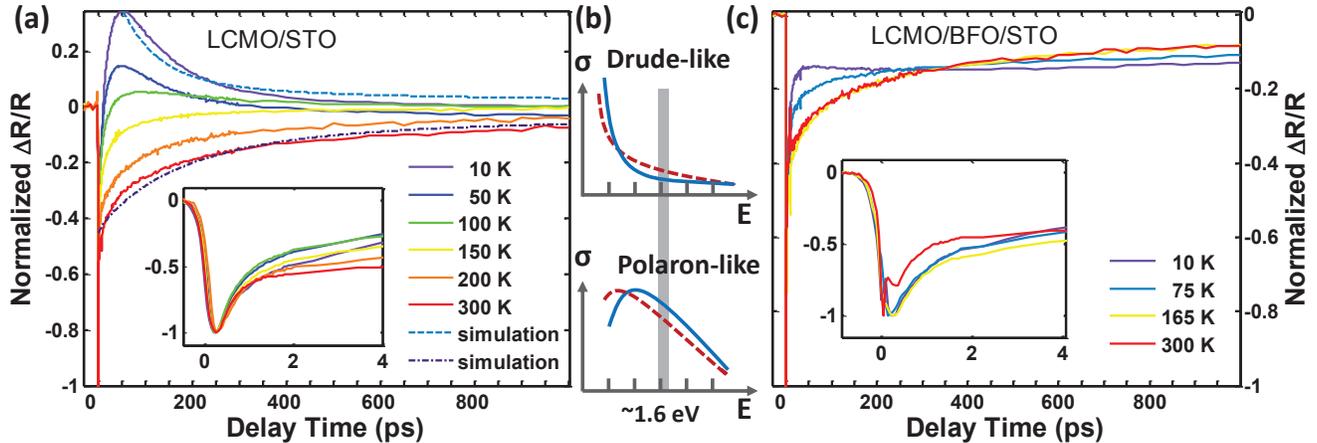}
\caption{ \label{f:DSWT} Photoinduced change in optical reflectivity as a function of time delay at various temperatures, measured at 1.59 eV in (a) a LCMO film and (c) a LCMO/BFO heterostructure.  In LCMO, $T_c$$\sim$160 K, consistent with that obtained through magnetotransport measurements. The dashed and dashed-dotted lines are numerical simulations of 1D thermal diffusion across the LCMO/STO interface at 10 K and 300 K, calculated as in Ref. \citenum{Sheu2011SSC}. Insets show the dynamics at early times. The spike-like feature in the data of (c) at 300 K at $t$=0 is due to the coherent artifact \cite{Rohit2011Book}; its superposition with the transient reflectivity signal leads to the observed sub-ps oscillatory signal. (b) Schematic diagram of the optical conductivity versus energy ($<$2.5 eV) for typical optimally doped manganites in both Drude-like (low temperature) and polaron-like (high temperature) phases (simplified cartoon redrawn from data of Ref. \citenum{Okimoto1995PRL}). The solid blue and dashed red lines illustrate the optical conductivity before and after laser excitation. The polaron peak is observed around $\sim$1-1.6 eV in various manganites.}
\end{center}
\end{figure*}

Here, we used UOS to uncover the polaronic nature of carrier transport in a LCMO/BFO heterostructure, originating from spin canting in LCMO near the interface with AFM ordered BFO. This was revealed through the difference in DSWT between a LCMO film and a LCMO/BFO heterostructure at low temperatures, and further supported by the observed increase in conductivity when a B field was used to increase the spin alignment. These results can thus provide a new avenue to engineering multiferroic functionality \cite{Wu2010NM}, since the AFM ordered spins in BFO interact with FM spins in LCMO across the interface and are coupled to the FE polarization \cite{Zhao2006NM,Wu2013PRL}.

The samples studied here are a thin film of optimally doped LCMO (thickness $t$=10 nm), which is an FM metal (FMM) below $T_c$ and a paramagnetic insulator (PMI) above $T_c$ ($\sim$260 K in the bulk, $\sim$160 K in our film), and a LCMO(10 nm)/BFO(50 nm) bilayer heterostructure, both grown on (001) SrTiO$_{3}$ (STO) substrates by pulsed laser (KrF) deposition. The heterostructure (LCMO/BFO) is deposited by switching the target without breaking the vacuum. The substrate temperature during film growth is initially optimized and maintained. The oxygen pressure during deposition is 100 mTorr. The samples are cooled to room temperature in pure oxygen (350 Torr) by turning off the power supply to the heater without further thermal treatment \cite{Lu2000Sample}. The degenerate optical pump-probe measurements use an amplified Ti:sapphire laser system producing pulses at a 250 kHz repetition rate with $\sim$150 fs duration and energy $\sim$4 $\mu$J/pulse at a center wavelength of 780 nm (1.59 eV). The incident pump fluence is $\sim$25 $\mu$J/cm$^{2}$, creating a photoexcited carrier density of $\sim$7$\times$10$^{18}$cm$^{-3}$ ($\sim$4$\times$10$^{-4}$ per unit cell). The photoinduced initial increase in the lattice temperature is 2-4 K, calculated from the heat capacity, which should decrease with increasing equilibrium
temperature. The B-field-dependent measurements are performed in a $\sim$5 Tesla (T) superconducting magnet.

To unravel quasiparticle dynamics in our LCMO/BFO heterostructure, it was essential to first understand the dynamics in our thin LCMO film. Previous UOS experiments on LCMO have shown that in thick films ($t>$75 nm), the first time constant ($\tau_{1}<$1 ps) arises from electron-phonon coupling in both FMM and PMI phases. The second time constant originates from spin-lattice relaxation ($\tau_{2}\sim$30-100 ps), slowing down near $T_c$ due to an increase in the spin specific heat, and the third time constant, $\tau_{3}$, is due to heat diffusion ($\tau_{3}>$1 ns) \cite{Lobad2001PRB,Ogasawar2003PRB,Bielecki2010PRB}. In our 10 nm LCMO film, temperature-dependent measurements of the transient photoinduced change in reflectivity, $\Delta$R/R (Fig. \ref{f:DSWT}(a)), revealed that $\tau_2$ and $\tau_3$ are much faster than in thicker films. However, thickness-dependent measurements on a series of LCMO films (not shown) revealed that their origin is the same \cite{ThicknessNote}. This was confirmed by a numerical simulation for 1D thermal diffusion across an interface, which demonstrated that the faster decay times in thin films result from the strong influence of the substrate thermal diffusion on thermal transport, rather than diffusion in the films themselves (dashed and dashed-dotted lines in Fig. 1(a)) \cite{Sheu2011SSC}. Therefore, as in thicker films, the $\Delta$R/R signals measured on our thin LCMO film originate from laser heating-induced DSWT \cite{Averitt2001PRL,Lobad2001PRB}. Importantly, these signals, obtained in a non-contact manner, can indicate whether the equilibrium state of LCMO is metallic or insulating, as follows.

In optimally doped manganites, the time-integrated optical conductivity displays a Drude-like feature in the low temperature FMM state  \cite{Kim1998PRL,Quijada1998PRB} (blue solid line in upper panel of Fig. \ref{f:DSWT}(b)). This feature evolves into a higher energy ($\sim$1-1.6 eV) peak as $T$ increases \cite{Kim1998PRL,Quijada1998PRB} (blue solid line in lower panel of Fig. \ref{f:DSWT}(b)). This SWT dominates the low energy physics ($<$3 eV) in manganites, and has been attributed to the appearance of Jahn-Teller (J-T) polarons as $T$ approaches $T_c$ from below, which trap electrons hopping from Mn$^{3+}$ to Mn$^{4+}$ sites, reducing the conductivity \cite{Kaplan1996PRL, Kim1998PRL, Quijada1998PRB, Hartinger2004PRB, Millis1996PRL, Millis1996PRB}. The formation of this polaron peak is considered to be a signature of competition between the electron kinetic energy and the J-T lattice distortion \cite{Millis1996PRB}. However, this temperature-induced SWT can be reversed by an applied B field. The applied field enhances the spin alignment, increasing the conductivity, which reduces the polaronic peak and recovers the Drude-like feature (i.e. following the blue solid line from the lower to the upper panel in Fig. \ref{f:DSWT}(b)) \cite{Tomioka1996PRB,Okimoto1998PRB,Tokura2000Science,Jung2000PRB,Millis1996PRB,Millis1996PRL,Millis1998Nature}.

DSWT then occurs after femtosecond photoexcitation, which transfers energy from the electronic subsystem to the lattice within a few ps, causing $\Delta$R/R at $\sim$1-1.6 eV to transiently reproduce the change in optical conductivity on this timescale \cite{Averitt2002JPCM,Okimoto2007JPSJ}. Therefore, below $T_c$, $\Delta$R/R$>$0 indicates a photoinduced increase in the lattice temperature, increasing the resistivity $\rho$ of the metallic state ($d\rho/dT >$0; the DSWT transiently redistributes the spectrum to higher energies (upper panel of Fig. \ref{f:DSWT}(b)), so $d$R/$dT$(1.59 eV)$>$0). Above $T_c$, $\Delta$R/R$<$0 symbolizes a photoinduced increase in the conductivity of the insulating state by liberating polarons ($d\rho/dT<$0; the DSWT transiently redistributes the spectrum to lower energies (lower panel of Fig. \ref{f:DSWT}(b)), so $d$R/$dT$(1.59 eV)$<$0). Representative 'resistive' ($T<T_c$) and 'conductive' ($T>T_c$) transients are given by the 10 K and 300 K traces in Fig. \ref{f:DSWT}(a), respectively. For clarity, we emphasize that a 'resistive' $\Delta$R/R transient indicates that the equilibrium state is metallic, while a 'conductive' transient indicates that the equilibrium state is insulating (due to polaronic transport). Finally, since an applied B field increases the conductivity, the 'conductive' transient measured in the insulating state evolves towards a 'resistive' transient as the field increases \cite{Lobad2001PRB,Okimoto2007JPSJ,Tobey2008PRL}. In this manner, we can use UOS to reveal whether LCMO is insulating or metallic in equilibrium. (See appendix A for more details on DSWT in manganites.)

We can now use this detailed characterization of our thin LCMO films as a basis for understanding carrier transport in the LCMO/BFO heterostructure. First, we note that the $\Delta$R/R signal in LCMO/BFO must be due to photoinduced changes in the LCMO layer, since the band gap of BFO is greater than 2.6 eV, preventing it from being directly photoexcited with our 1.59 eV pump photons \cite{Sheu2012APL,Wen2013PRL}. In addition, we expect that the temperature-dependent DSWT in optimally doped LCMO should stay the same when changing the material underneath, as long as no novel interface state is formed.

This is observed in LCMO/BFO above $\sim$165 K (through comparing Fig. \ref{f:DSWT}(a) and (c)), indicating that the relaxation mechanisms remain the same at high temperatures.  However, at low temperatures ($T<$100 K) the resistive transient observed in LCMO alone is replaced by a conductive transient in LCMO/BFO, implying a change in the magnetic and/or metallic properties of LCMO when grown on BFO. In particular, the observed conductive transient at low $T$, similar to that observed at high $T$ in LCMO alone, suggests that LCMO is more insulating when grown on BFO.
	
This is likely most significant near the LCMO/BFO interface, where the two materials can directly interact \cite{Yu2010PRL,Yu2012MatToday,Wu2010NM,Wu2013PRL}. Our UOS measurements can sensitively probe this interface. Since our 10 nm LCMO film is much thinner than the laser absorption depth ($\sim$100 nm), a substantial portion of the 1.59 eV photons reach the LCMO/BFO interface, such that any photoinduced changes in the interfacial states could significantly influence $\Delta$R/R \cite{multiRNote}.

\begin{figure}[tb]
\begin{center}
\includegraphics[width=3.2in]{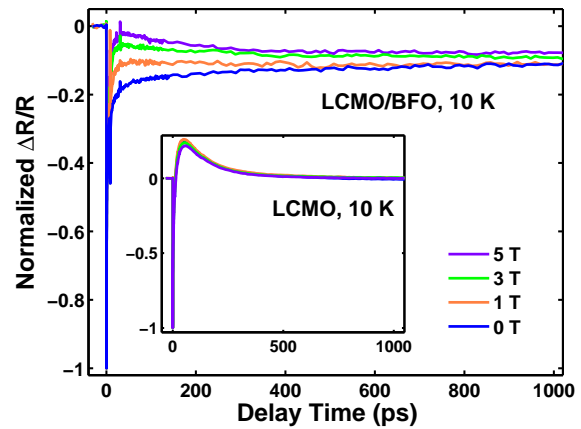}
\caption{ \label{f:Bfield}
Photoinduced time-dependent reflectivity change in LCMO/BFO as a function of B field at 10 K. The inset shows the $\Delta$R/R signal from a single LCMO film, taken under the same experimental conditions. }
\end{center}
\end{figure}

To gain more insight into the insulating nature of the LCMO/BFO interface, we performed low temperature measurements in a B field up to 5 T directed along the surface normal, allowing us to directly rotate spins (and thus overcome interfacial spin pinning \cite{Wu2013PRL}).  Fig. \ref{f:Bfield} displays the measured $\Delta$R/R signal for LCMO/BFO as a function of B field at 10 K, where both samples were zero-field cooled. The inset shows the signal from the individual LCMO film, measured under the same conditions as LCMO/BFO, which reveals no discernible change since the magnetization is saturated in the low temperature FMM state. However, in the heterostructure, we find a large field-tunable DSWT. As the B field increases, the conductive transient looks more and more resistive, suggesting growth of the metallic state \cite{Millis1996PRB,Okimoto1998PRB}.

Overall, our femtosecond optical data suggests that interfacial LCMO/BFO is insulating in equilibrium, and that the application of a B field causes it to become more conductive. Because both samples are zero-field cooled, we can rule out the possibility of domains as the origin of this behavior in LCMO/BFO. Another possibility, charge transfer between LCMO and BFO, is prohibited in equilibrium, as Ref. \citenum{Yu2010PRL} measures no change in the valence state of Fe$^{3+}$. Photostriction has previously been observed in BFO \cite{Kundys2010NM}, but is unlikely to affect our results, since both our previous experiments \cite{Sheu2012APL} and time-resolved x-ray diffraction \cite{Wen2013PRL} on BFO measured no response upon $\sim$1.5 eV excitation. Finally, the difference in thermal diffusion between BFO/STO and STO (a factor of $\sim$5) is also unlikely to influence our results, since the high temperature data ($\geq$165 K) on both samples is very similar (Fig. \ref{f:DSWT}), and thermal diffusion would not be affected by a B field. A more detailed exclusion of these effects is presented in Appendix B. Instead, the conductive transient observed at low temperatures and high B fields in LCMO/BFO suggests that polarons play a significant role in the observed phenomena (similar to the PMI phase of LCMO, Fig. \ref{f:DSWT}(a)), likely by decreasing the charge kinetic energy near the LCMO/BFO interface.  Such polaronic behavior could arise from AFM order, orbital order or a lattice distortion that creates a trapping potential, all of which are likely coupled.
	
\begin{figure}[tb]
\begin{center}
\includegraphics[width=3.3in]{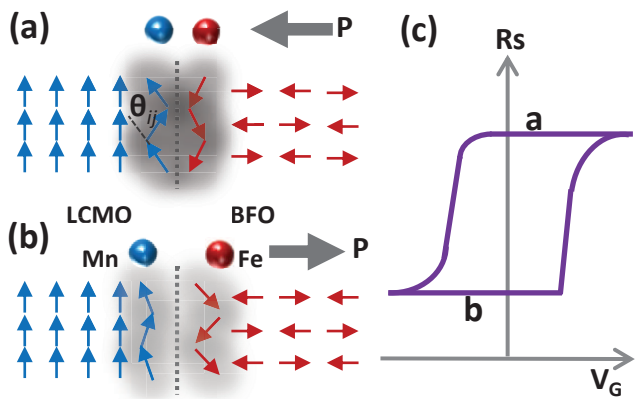}
\caption{ \label{f:polaron} Schematic diagrams showing spin tuning through ferroelectric switching, where we only show the spin component perpendicular to the interface in BFO \cite{SpinNote}. The spin alignment is depicted in LCMO(FM)/BFO(G-type AFM) with FE polarization (a) towards the interface (darker gray with stronger interaction) and (b) away from the interface (lighter gray with weaker interaction). (c) The corresponding magnetoresistance $R_s$ in states (a) and (b) as a function of gate voltage $V_g$, sketched from Ref. \citenum{Wu2010NM}. }
\end{center}
\end{figure}

The most likely explanation for the observed polaronic behavior when LCMO was grown on BFO is an induced AFM order in LCMO (reducing the FM order). This can result from interfacial coupling to G-type BFO, with its alternating spin alignment of Fe$^{3+}$ ions \cite{Yu2010PRL} at the interface, through the interfacial spin-spin interaction $J_{\text{FM-AFM(G)}}$ (Fig. \ref{f:polaron}(a)). For more insight, we consider that in optimally doped manganites, transport is governed by the double exchange interaction, in which the charge kinetic energy H$_k$ is determined by the hopping amplitude $t_{ij}$(1+$<\textbf{S}_{i}\textbf{S}_{j}>$/S$^{2})^{1/2}$=$t_{ij}$cos($\theta_{ij}/2$) \cite{Millis1996PRB}, where $\textbf{S}_i,\textbf{S}_j$ are nearest neighbor Mn $t_{2g}$ spins and $\theta_{ij}$ is the angle between them (Fig. \ref{f:polaron}(a)). In the general Hamiltonian for Mn $e_{g}$ electrons, hopping is linked to phonons (H$_{ph}$) and the electron-phonon coupling (H$_{e-p}$), both of which are related to the formation of polarons \cite{Millis1996PRB}. As H$_{k}$ surpasses H$_{ph}$ and H$_{e-p}$, the spins become more ordered and the state becomes more metallic, i.e. with increased conductivity. In LCMO/BFO, the induced AFM order at the interface will increase $\theta_{ij}$ in LCMO (Fig \ref{f:polaron}(a)), suppressing in-plane electron hopping. This results in a discernible difference between LCMO and LCMO/BFO in a high B field, which increases the alignment of the induced AFM spins \cite{Okimoto1998PRB,Tokura2000Science} while keeping the ordered FM spins saturated. We note that neutron diffraction measurements on LSMO/BFO superlattices grown in the same system at LANL have revealed a reduced magnetization in the LSMO layers, supporting this hypothesis \cite{Singh2013unpublished}.

This picture, based on an induced interfacial AFM order in LCMO, can now be used to shed light on previous work in which the magnetotransport was tuned by switching the FE polarization \cite{Wu2010NM,Wu2013PRL}. In LCMO/BFO, the separation $l$ between Fe$^{3+}$ and Mn$^{3+/4+}$ ions across the interface affects the strength of $J_{\text{FM-AFM(G)}}$ and therefore $\theta_{ij}$ (i.e. $J_{\text{FM-AFM(G)}}$ exponentially decays as $\sim e^{-l/d}$ \cite{SwitchNote}, where $d$ is the atomic distance, $\sim$3-5 \AA).   Thus, when the BFO film is electrically poled such that the polarization points to the interface (displacing Fe$^{3+}$ ions towards the interface, Fig. \ref{f:polaron}(a)), the reduction in $l$ enhances the effective 'AFM field' (the magnitude of $J_{\text{FM-AFM(G)}}$ and $\theta_{ij}$ increase), decreasing the hopping amplitude. When the FE polarity is switched, the increase in $l$ (Fig. \ref{f:polaron}(b)) causes the magnitude of $J_{\text{FM-AFM(G)}}$ to decrease, increasing the hopping amplitude. The change in $l$ between the two FE polarity states is $>\sim$0.7 \AA~(the lower limit is estimated from as-grown, non-electrically poled LSMO/BFO \cite{Chang2011AM}), which is significant compared to $d$. Therefore, tunable magnetotransport can be observed when switching the FE polarization in BFO \cite{Wu2010NM,Yu2012MatToday} (Fig. \ref{f:polaron}(c)).

Our results thus indicate that polarons originate from competing interfacial FM/AFM order in LCMO/BFO. This mutual spin interaction results in a measurable magnetic moment in the Fe ion of G-type BFO \cite{Yu2010PRL} and an induced AFM order in the Mn ion of FM LCMO. The ionic distance between these two transition metals can be then used to tune magnetotransport in these systems.

In summary, our UOS studies reveal a difference in the DSWT between single layer LCMO and a LCMO/BFO heterostructure. This indicates that interfacial transport in LCMO/BFO is polaronic in nature and is modified in a magnetic field, while no change is observed in LCMO. In the framework of the double exchange interaction, we suggest that an interfacial AFM order in LCMO is induced by an effective 'AFM field' arising from G-type antiferromagnetic ordered BFO, which reduces in-plane electron hopping.  Furthermore, by switching the FE polarity in BFO, one can change the coupling strength by tuning the separation between Fe-Mn, providing a new avenue for achieving tunable magnetotransport \cite{Wu2010NM,Yu2012MatToday,Wu2013PRL}.

\section{acknowledgment}

This work was performed at the Center for Integrated Nanotechnologies, a U.S. Department of Energy, Office of Basic Energy Sciences user facility and under the auspices of the Department of Energy, Office of Basic Energy Sciences, Division of Material Sciences. It was also partially supported by the NNSA's Laboratory Directed Research and Development Program. Los Alamos National Laboratory, an affirmative action equal opportunity employer, is operated by Los Alamos National Security, LLC, for the National Nuclear Security Administration of the U. S. Department of Energy under contract DE-AC52-06NA25396.

\section{Appendix A: Dynamic Spectral Weight Transfer Revealed by Degenerate Optical Pump-Probe Spectroscopy}

Colossal magnetoresistance in manganites originates from a giant change in the conductivity that is controlled by the spin alignment (either through applying an external magnetic field or going to low temperatures).  This in turn dominates the optical properties from terahertz to visible frequencies through "spectral weight transfer" (SWT), as schematically depicted in Fig. \ref{f:DSWT}(b) of the manuscript. SWT is a well established concept in manganites (described in more detail in Refs. \cite{Millis1996PRB,Okimoto1995PRL,Kim1998PRL,Quijada1998PRB,Kaplan1996PRL,Hartinger2004PRB,BasovReview2011}). At low temperatures, the optical conductivity in optimally doped manganites is dominated by a low frequency Drude peak, which is a signature of the metallic state. As the temperature increases, the amplitude of the Drude peak diminishes as the conductivity decreases (due to decreasing spin alignment), and a peak in the near-to-mid-infrared (IR) concurrently increases, which has been widely attributed to the trapping of free carriers into polaronic states \cite{Millis1996PRB,Kim1998PRL,Quijada1998PRB,Kaplan1996PRL,Hartinger2004PRB}. In the high temperature insulating phase (above $T_C$), a strong external magnetic (B) field can induce SWT from the IR polaronic peak to the low frequency Drude peak by orienting spins (and thus increasing the conductivity). In the low temperature metallic phase, the B field will not increase the conductivity, since the spins are already well aligned.

Optical measurements can therefore be significantly influenced by the spin alignment through SWT. This can then be extended to non-equilibrium optical measurements, in which the femtosecond pump pulse introduces heat when the sample is at low temperatures (resulting in SWT to higher frequencies) and liberates carriers trapped in polaronic states when the sample is at high temperatures (initiating SWT to lower frequencies); this is the idea of "dynamic spectral weight transfer" (DSWT) described in the main manuscript (and schematically depicted in Fig. \ref{f:DSWT}(b)).  DSWT is a well-known concept that has been widely shown to dominate the response in ultrafast optical measurements on manganites \cite{Averitt2001PRL,Lobad2001PRB,Tobey2008PRL,Okimoto2007JPSJ,Ogasawar2003PRB}.

Further support for this concept comes from previous work on similar samples: in Figure 14 of Ref. \cite{Averitt2002JPCM}, it was clearly shown that for an LCMO film, the time constant associated with DSWT was nearly identical when probing at both 1.5 eV and THz frequencies. This is strong evidence that a 1.5 eV probe is sensitive to DSWT, even in the absence of complementary optical-pump, THz-probe measurements. Therefore, the degenerate 1.5 eV pump-probe measurements described here can clearly reveal the transition from Drude-like to polaronic behavior in LCMO through DSWT, since the origin of the photoinduced change in optical properties is the same.

Finally, we emphasize that the strongest evidence for the sensitivity of our ultrafast optical reflectivity measurements to the spin alignment is given by the field tuning experiment displayed in Fig. \ref{f:polaron} of our manuscript: while we saw significant field tuning in LCMO/BFO, we observed no change in LCMO, strongly supporting the field-induced increase in the conductivity of LCMO/BFO, while the conductivity of LCMO/STO was saturated as all the spins were already aligned at 10 K.

\section{Appendix B: Exclusion of other possible contributions to our data}

We have considered the influence of lattice mismatch and/or modifications of the lattice properties due to the change in the layer underneath LCMO. Therefore, we have performed temperature-dependent ultrafast optical measurements of LC(S)MO on MgO (on which manganites are relaxed, as measured with x-ray diffraction) and on STO (on which manganites are strained, also measured with x-ray diffraction). We found that the results were very similar for different substrates, indicating that simple lattice mismatch or random interfacial disorder should not influence the observed data. We also considered the effect of changes in the thermal diffusivity on our data. However, we can exclude this possibility, since our data on LCMO/BFO/STO and LCMO/STO is very similar at higher temperatures (Fig. \ref{f:DSWT}(a) and \ref{f:DSWT}(c)), despite the fact that the thermal conductivities of BFO ($\sim$2 W/m/K) and STO (11 W/m/K) and the heat capacities of BFO ($\sim$2 J/cm$^3$/K) and STO ($\sim$2.2 J/cm$^3$/K) at 300 K lead to a factor of $\sim$5 difference in the diffusivities of BFO ($\sim$0.01 cm$^{2}$/s) and STO ($\sim$0.05 cm$^2$/s). This should lead to a corresponding difference in the time constant associated with thermal decay in our data, but we observed no significant difference (traces at 300 K in Fig. \ref{f:DSWT}(a) and \ref{f:DSWT}(c)). Furthermore, the difference in thermal diffusivity increases with decreasing temperature \cite{Hopkins2013APL} while the specific heat of BFO varies little and that of STO decreases 30$\%$ between 165 and 300 K, but our LCMO/BFO/STO and LCMO/STO data remains quite similar down to $\sim$165 K, making it very unlikely that this would significantly influence our results (particularly since the 50 nm BFO layer is much thinner than the STO substrate, which thus serves as a heat sink). Finally, because an AFM material has no response to an external B field of the magnitude used in our experiments, if the thermal diffusivity of BFO is the origin of the observed difference, then we would not expect to see the B field dependence shown in Fig. \ref{f:Bfield}.

It is also worth considering the possible influence of photostrictive effects on our data, as they have been previously observed in BFO \cite{Kundys2010NM}. In this context, it is important to emphasize that our optical pump pulses only photoexcite the LCMO layer, since the pump photon energy (1.59 eV) is much lower than the bandgap ($\sim$2.6 eV) of BFO. We have extensively studied both single crystal BFO and thin films of BFO grown on STO, and we have never found an observable optical response with a 1.59 eV pump, even at much higher fluences when two-photon processes could play a role (we have tried fluences up to 2 orders of magnitude larger than we used in this manuscript to pump BFO, with no observable optical response).  Our detailed studies of ultrafast dynamics in single crystal BFO can be found in Ref. \cite{Sheu2012APL}, and comparison of the dynamics between BFO single crystals and films is available in Ref. \cite{Sheu2013EPJ}. In addition, if a photostrictive effect arising from the BFO layer somehow influenced our data, we would expect to see a significant difference between the dynamics in LCMO/BFO/STO and LCMO/STO at all temperatures, which is not the case at high temperatures (Fig. \ref{f:DSWT}). Therefore, within our experimental sensitivity, we conclude that BFO is not directly excited through either one-photon or multiphoton processes, and can therefore exclude the possibility of structural changes due to direct excitation of BFO. Additional support can be found in Ref. \cite{Wen2013PRL}, where photoinduced changes in the lattice constants of a BFO film were discovered using time-resolved x-ray diffraction with above gap excitation, but not observed when using a 1.55 eV pump.

Finally, we note that other methods used to unravel quasiparticle dynamics in oxides, such as time-resolved second harmonic generation (SHG) and magneto-optical Kerr effect (MOKE) are not as suitable for studying polaronic behavior in LCMO/BFO. Our preliminary
SHG measurements on LCMO/BFO indicate that BFO is the origin of the observed SHG signal, due to broken inversion symmetry from its FE order. This is therefore insensitive to the interfacial spin order and conductivity of the LCMO layer. Furthermore, the MOKE signal would be influenced by both the signal from FM order throughout the LCMO layer as well as the induced FM order in the BFO layer, making it difficult to isolate the dynamics in the LCMO layer near the interface. Therefore, ultrafast optical reflection or transmission measurements, at an energy that is not sensitive to the optical properties of BFO, are actually more useful and straightforward to interpret in this case.


\begin{thebibliography}{60}
\expandafter\ifx\csname natexlab\endcsname\relax\def\natexlab#1{#1}\fi
\expandafter\ifx\csname bibnamefont\endcsname\relax
  \def\bibnamefont#1{#1}\fi
\expandafter\ifx\csname bibfnamefont\endcsname\relax
  \def\bibfnamefont#1{#1}\fi
\expandafter\ifx\csname citenamefont\endcsname\relax
  \def\citenamefont#1{#1}\fi
\expandafter\ifx\csname url\endcsname\relax
  \def\url#1{\texttt{#1}}\fi
\expandafter\ifx\csname urlprefix\endcsname\relax\def\urlprefix{URL }\fi
\providecommand{\bibinfo}[2]{#2}
\providecommand{\eprint}[2][]{\url{#2}}

\bibitem[{\citenamefont{Eerenstein et~al.}(2007)\citenamefont{Eerenstein,
  Wiora, Prieto, Scott, and Mathur}}]{Eerenstein2007NM}
\bibinfo{author}{\bibfnamefont{W.}~\bibnamefont{Eerenstein}},
  \bibinfo{author}{\bibfnamefont{M.}~\bibnamefont{Wiora}},
  \bibinfo{author}{\bibfnamefont{J.~L.} \bibnamefont{Prieto}},
  \bibinfo{author}{\bibfnamefont{J.~F.} \bibnamefont{Scott}}, \bibnamefont{and}
  \bibinfo{author}{\bibfnamefont{N.~D.} \bibnamefont{Mathur}},
  \bibinfo{journal}{Nat. Mater.} \textbf{\bibinfo{volume}{6}},
  \bibinfo{pages}{348} (\bibinfo{year}{2007}).

\bibitem[{\citenamefont{Wu et~al.}(2010)\citenamefont{Wu, Cybart, Yu, Rossell,
  Zhang, Ramesh, and Dynes}}]{Wu2010NM}
\bibinfo{author}{\bibfnamefont{S.~M.} \bibnamefont{Wu}},
  \bibinfo{author}{\bibfnamefont{S.~A.} \bibnamefont{Cybart}},
  \bibinfo{author}{\bibfnamefont{P.}~\bibnamefont{Yu}},
  \bibinfo{author}{\bibfnamefont{M.~D.} \bibnamefont{Rossell}},
  \bibinfo{author}{\bibfnamefont{J.~X.} \bibnamefont{Zhang}},
  \bibinfo{author}{\bibfnamefont{R.}~\bibnamefont{Ramesh}}, \bibnamefont{and}
  \bibinfo{author}{\bibfnamefont{R.~C.} \bibnamefont{Dynes}},
  \bibinfo{journal}{Nat. Mater.} \textbf{\bibinfo{volume}{9}},
  \bibinfo{pages}{756} (\bibinfo{year}{2010}).

\bibitem[{\citenamefont{Yu et~al.}(2012)\citenamefont{Yu, Chu, and
  Ramesh}}]{Yu2012MatToday}
\bibinfo{author}{\bibfnamefont{P.}~\bibnamefont{Yu}},
  \bibinfo{author}{\bibfnamefont{Y.-H.} \bibnamefont{Chu}}, \bibnamefont{and}
  \bibinfo{author}{\bibfnamefont{R.}~\bibnamefont{Ramesh}},
  \bibinfo{journal}{Mat. Today} \textbf{\bibinfo{volume}{15}},
  \bibinfo{pages}{320} (\bibinfo{year}{2012}).

\bibitem[{\citenamefont{Heber}(2009)}]{Heber2009Nature}
\bibinfo{author}{\bibfnamefont{J.}~\bibnamefont{Heber}},
  \bibinfo{journal}{Nature} \textbf{\bibinfo{volume}{459}}, \bibinfo{pages}{28}
  (\bibinfo{year}{2009}).

\bibitem[{\citenamefont{Hwang et~al.}(2012)\citenamefont{Hwang, Iwasa,
  Kawasaki, Keimer, Nagaosa, and Tokura}}]{Hwang2012NM}
\bibinfo{author}{\bibfnamefont{H.~Y.} \bibnamefont{Hwang}},
  \bibinfo{author}{\bibfnamefont{Y.}~\bibnamefont{Iwasa}},
  \bibinfo{author}{\bibfnamefont{M.}~\bibnamefont{Kawasaki}},
  \bibinfo{author}{\bibfnamefont{B.}~\bibnamefont{Keimer}},
  \bibinfo{author}{\bibfnamefont{N.}~\bibnamefont{Nagaosa}}, \bibnamefont{and}
  \bibinfo{author}{\bibfnamefont{Y.}~\bibnamefont{Tokura}},
  \bibinfo{journal}{Nat. Mater.} \textbf{\bibinfo{volume}{11}},
  \bibinfo{pages}{103} (\bibinfo{year}{2012}).

\bibitem[{\citenamefont{Ueda et~al.}(1998)\citenamefont{Ueda, Tabata, and
  Kawai}}]{Ueda1998Science}
\bibinfo{author}{\bibfnamefont{K.}~\bibnamefont{Ueda}},
  \bibinfo{author}{\bibfnamefont{H.}~\bibnamefont{Tabata}}, \bibnamefont{and}
  \bibinfo{author}{\bibfnamefont{T.}~\bibnamefont{Kawai}},
  \bibinfo{journal}{Science} \textbf{\bibinfo{volume}{280}},
  \bibinfo{pages}{1064} (\bibinfo{year}{1998}).

\bibitem[{\citenamefont{Ohtomo and Hwang}(2004)}]{Ohtomo2004Nature}
\bibinfo{author}{\bibfnamefont{A.}~\bibnamefont{Ohtomo}} \bibnamefont{and}
  \bibinfo{author}{\bibfnamefont{H.~Y.} \bibnamefont{Hwang}},
  \bibinfo{journal}{Nature} \textbf{\bibinfo{volume}{427}},
  \bibinfo{pages}{423} (\bibinfo{year}{2004}).

\bibitem[{\citenamefont{Chakhalian et~al.}(2007)\citenamefont{Chakhalian,
  Freeland, Habermeier, Cristiani, Khaliullin, van Veenendaal, and
  Keimer}}]{Chakhalian2007Science}
\bibinfo{author}{\bibfnamefont{J.}~\bibnamefont{Chakhalian}},
  \bibinfo{author}{\bibfnamefont{J.~W.} \bibnamefont{Freeland}},
  \bibinfo{author}{\bibfnamefont{H.-U.} \bibnamefont{Habermeier}},
  \bibinfo{author}{\bibfnamefont{G.}~\bibnamefont{Cristiani}},
  \bibinfo{author}{\bibfnamefont{G.}~\bibnamefont{Khaliullin}},
  \bibinfo{author}{\bibfnamefont{M.}~\bibnamefont{van Veenendaal}},
  \bibnamefont{and} \bibinfo{author}{\bibfnamefont{B.}~\bibnamefont{Keimer}},
  \bibinfo{journal}{Science} \textbf{\bibinfo{volume}{318}},
  \bibinfo{pages}{1114} (\bibinfo{year}{2007}).

\bibitem[{\citenamefont{B\'{e}a
  et~al.}(2006{\natexlab{a}})\citenamefont{B\'{e}a, Bibes, Sirena, Herranz,
  Bouzehouane, Jacquet, Fusil, Paruch, Dawber, Contour et~al.}}]{Bea2006APL}
\bibinfo{author}{\bibfnamefont{H.}~\bibnamefont{B\'{e}a}},
  \bibinfo{author}{\bibfnamefont{M.}~\bibnamefont{Bibes}},
  \bibinfo{author}{\bibfnamefont{M.}~\bibnamefont{Sirena}},
  \bibinfo{author}{\bibfnamefont{G.}~\bibnamefont{Herranz}},
  \bibinfo{author}{\bibfnamefont{K.}~\bibnamefont{Bouzehouane}},
  \bibinfo{author}{\bibfnamefont{E.}~\bibnamefont{Jacquet}},
  \bibinfo{author}{\bibfnamefont{S.}~\bibnamefont{Fusil}},
  \bibinfo{author}{\bibfnamefont{P.}~\bibnamefont{Paruch}},
  \bibinfo{author}{\bibfnamefont{M.}~\bibnamefont{Dawber}},
  \bibinfo{author}{\bibfnamefont{J.-P.} \bibnamefont{Contour}},
  \bibnamefont{et~al.}, \bibinfo{journal}{Appl. Phys. Lett.}
  \textbf{\bibinfo{volume}{88}}, \bibinfo{pages}{062502}
  (\bibinfo{year}{2006}{\natexlab{a}}).

\bibitem[{\citenamefont{B\'{e}a
  et~al.}(2006{\natexlab{b}})\citenamefont{B\'{e}a, Bibes, Cherifi, Nolting,
  Warot-Fonrose, Fusil, Herranz, Deranlot, Jacquet, Bouzehouane
  et~al.}}]{Bea2006APLEB}
\bibinfo{author}{\bibfnamefont{H.}~\bibnamefont{B\'{e}a}},
  \bibinfo{author}{\bibfnamefont{M.}~\bibnamefont{Bibes}},
  \bibinfo{author}{\bibfnamefont{S.}~\bibnamefont{Cherifi}},
  \bibinfo{author}{\bibfnamefont{F.}~\bibnamefont{Nolting}},
  \bibinfo{author}{\bibfnamefont{B.}~\bibnamefont{Warot-Fonrose}},
  \bibinfo{author}{\bibfnamefont{S.}~\bibnamefont{Fusil}},
  \bibinfo{author}{\bibfnamefont{G.}~\bibnamefont{Herranz}},
  \bibinfo{author}{\bibfnamefont{C.}~\bibnamefont{Deranlot}},
  \bibinfo{author}{\bibfnamefont{E.}~\bibnamefont{Jacquet}},
  \bibinfo{author}{\bibfnamefont{K.}~\bibnamefont{Bouzehouane}},
  \bibnamefont{et~al.}, \bibinfo{journal}{Appl. Phys. Lett.}
  \textbf{\bibinfo{volume}{89}}, \bibinfo{pages}{242114}
  (\bibinfo{year}{2006}{\natexlab{b}}).

\bibitem[{\citenamefont{Hambe et~al.}(2010)\citenamefont{Hambe, Petraru,
  Pertsev, Munroe, Nagarajan, and Kohlstedt}}]{Hambe2010AM}
\bibinfo{author}{\bibfnamefont{M.}~\bibnamefont{Hambe}},
  \bibinfo{author}{\bibfnamefont{A.}~\bibnamefont{Petraru}},
  \bibinfo{author}{\bibfnamefont{N.~A.} \bibnamefont{Pertsev}},
  \bibinfo{author}{\bibfnamefont{P.}~\bibnamefont{Munroe}},
  \bibinfo{author}{\bibfnamefont{V.}~\bibnamefont{Nagarajan}},
  \bibnamefont{and}
  \bibinfo{author}{\bibfnamefont{H.}~\bibnamefont{Kohlstedt}},
  \bibinfo{journal}{Adv. Funct. Mater.} \textbf{\bibinfo{volume}{20}},
  \bibinfo{pages}{2436} (\bibinfo{year}{2010}).

\bibitem[{\citenamefont{Borisevich et~al.}(2010)\citenamefont{Borisevich,
  Chang, Huijben, Oxley, Okamoto, Niranjan, Burton, Tsymbal, Chu, Yu
  et~al.}}]{Borisevich2010PRL}
\bibinfo{author}{\bibfnamefont{A.~Y.} \bibnamefont{Borisevich}},
  \bibinfo{author}{\bibfnamefont{H.~J.} \bibnamefont{Chang}},
  \bibinfo{author}{\bibfnamefont{M.}~\bibnamefont{Huijben}},
  \bibinfo{author}{\bibfnamefont{M.~P.} \bibnamefont{Oxley}},
  \bibinfo{author}{\bibfnamefont{S.}~\bibnamefont{Okamoto}},
  \bibinfo{author}{\bibfnamefont{M.~K.} \bibnamefont{Niranjan}},
  \bibinfo{author}{\bibfnamefont{J.~D.} \bibnamefont{Burton}},
  \bibinfo{author}{\bibfnamefont{E.~Y.} \bibnamefont{Tsymbal}},
  \bibinfo{author}{\bibfnamefont{Y.~H.} \bibnamefont{Chu}},
  \bibinfo{author}{\bibfnamefont{P.}~\bibnamefont{Yu}}, \bibnamefont{et~al.},
  \bibinfo{journal}{Phys. Rev. Lett.} \textbf{\bibinfo{volume}{105}},
  \bibinfo{pages}{087204} (\bibinfo{year}{2010}).

\bibitem[{\citenamefont{Zhao et~al.}(2006)\citenamefont{Zhao, Scholl,
  Zavaliche, Lee, Barry, Doran, Cruz, Chu, Ederer, Spaldin
  et~al.}}]{Zhao2006NM}
\bibinfo{author}{\bibfnamefont{T.}~\bibnamefont{Zhao}},
  \bibinfo{author}{\bibfnamefont{A.}~\bibnamefont{Scholl}},
  \bibinfo{author}{\bibfnamefont{F.}~\bibnamefont{Zavaliche}},
  \bibinfo{author}{\bibfnamefont{K.}~\bibnamefont{Lee}},
  \bibinfo{author}{\bibfnamefont{M.}~\bibnamefont{Barry}},
  \bibinfo{author}{\bibfnamefont{A.}~\bibnamefont{Doran}},
  \bibinfo{author}{\bibfnamefont{M.~P.} \bibnamefont{Cruz}},
  \bibinfo{author}{\bibfnamefont{Y.~H.} \bibnamefont{Chu}},
  \bibinfo{author}{\bibfnamefont{C.}~\bibnamefont{Ederer}},
  \bibinfo{author}{\bibfnamefont{N.~A.} \bibnamefont{Spaldin}},
  \bibnamefont{et~al.}, \bibinfo{journal}{Nat. Mater.}
  \textbf{\bibinfo{volume}{5}}, \bibinfo{pages}{823} (\bibinfo{year}{2006}).

\bibitem[{\citenamefont{Yu et~al.}(2010)\citenamefont{Yu, Lee, Okamoto,
  Rossell, Huijben, Yang, He, Zhang, Yang, Lee et~al.}}]{Yu2010PRL}
\bibinfo{author}{\bibfnamefont{P.}~\bibnamefont{Yu}},
  \bibinfo{author}{\bibfnamefont{J.-S.} \bibnamefont{Lee}},
  \bibinfo{author}{\bibfnamefont{S.}~\bibnamefont{Okamoto}},
  \bibinfo{author}{\bibfnamefont{M.~D.} \bibnamefont{Rossell}},
  \bibinfo{author}{\bibfnamefont{M.}~\bibnamefont{Huijben}},
  \bibinfo{author}{\bibfnamefont{C.-H.} \bibnamefont{Yang}},
  \bibinfo{author}{\bibfnamefont{Q.}~\bibnamefont{He}},
  \bibinfo{author}{\bibfnamefont{J.~X.} \bibnamefont{Zhang}},
  \bibinfo{author}{\bibfnamefont{S.~Y.} \bibnamefont{Yang}},
  \bibinfo{author}{\bibfnamefont{M.~J.} \bibnamefont{Lee}},
  \bibnamefont{et~al.}, \bibinfo{journal}{Phys. Rev. Lett.}
  \textbf{\bibinfo{volume}{105}}, \bibinfo{pages}{027201}
  (\bibinfo{year}{2010}).

\bibitem[{\citenamefont{Calder\'on et~al.}(2011)\citenamefont{Calder\'on,
  Liang, Yu, Salafranca, Dong, Yunoki, Brey, Moreo, and
  Dagotto}}]{Calderon2011PRB}
\bibinfo{author}{\bibfnamefont{M.~J.} \bibnamefont{Calder\'on}},
  \bibinfo{author}{\bibfnamefont{S.}~\bibnamefont{Liang}},
  \bibinfo{author}{\bibfnamefont{R.}~\bibnamefont{Yu}},
  \bibinfo{author}{\bibfnamefont{J.}~\bibnamefont{Salafranca}},
  \bibinfo{author}{\bibfnamefont{S.}~\bibnamefont{Dong}},
  \bibinfo{author}{\bibfnamefont{S.}~\bibnamefont{Yunoki}},
  \bibinfo{author}{\bibfnamefont{L.}~\bibnamefont{Brey}},
  \bibinfo{author}{\bibfnamefont{A.}~\bibnamefont{Moreo}}, \bibnamefont{and}
  \bibinfo{author}{\bibfnamefont{E.}~\bibnamefont{Dagotto}},
  \bibinfo{journal}{Phys. Rev. B} \textbf{\bibinfo{volume}{84}},
  \bibinfo{pages}{024422} (\bibinfo{year}{2011}).

\bibitem[{\citenamefont{Wu et~al.}(2013)\citenamefont{Wu, Cybart, Yi, Parker,
  Ramesh, and Dynes}}]{Wu2013PRL}
\bibinfo{author}{\bibfnamefont{S.~M.} \bibnamefont{Wu}},
  \bibinfo{author}{\bibfnamefont{S.~A.} \bibnamefont{Cybart}},
  \bibinfo{author}{\bibfnamefont{D.}~\bibnamefont{Yi}},
  \bibinfo{author}{\bibfnamefont{J.~M.} \bibnamefont{Parker}},
  \bibinfo{author}{\bibfnamefont{R.}~\bibnamefont{Ramesh}}, \bibnamefont{and}
  \bibinfo{author}{\bibfnamefont{R.~C.} \bibnamefont{Dynes}},
  \bibinfo{journal}{Phys. Rev. Lett.} \textbf{\bibinfo{volume}{110}},
  \bibinfo{pages}{067202} (\bibinfo{year}{2013}).

\bibitem[{\citenamefont{Averitt et~al.}(2001)\citenamefont{Averitt, Lobad,
  Kwon, Trugman, Thorsm\o{}lle, and Taylor}}]{Averitt2001PRL}
\bibinfo{author}{\bibfnamefont{R.~D.} \bibnamefont{Averitt}},
  \bibinfo{author}{\bibfnamefont{A.~I.} \bibnamefont{Lobad}},
  \bibinfo{author}{\bibfnamefont{C.}~\bibnamefont{Kwon}},
  \bibinfo{author}{\bibfnamefont{S.~A.} \bibnamefont{Trugman}},
  \bibinfo{author}{\bibfnamefont{V.~K.} \bibnamefont{Thorsm\o{}lle}},
  \bibnamefont{and} \bibinfo{author}{\bibfnamefont{A.~J.}
  \bibnamefont{Taylor}}, \bibinfo{journal}{Phys. Rev. Lett.}
  \textbf{\bibinfo{volume}{87}}, \bibinfo{pages}{017401}
  (\bibinfo{year}{2001}).

\bibitem[{\citenamefont{Lobad et~al.}(2001)\citenamefont{Lobad, Averitt, and
  Taylor}}]{Lobad2001PRB}
\bibinfo{author}{\bibfnamefont{A.~I.} \bibnamefont{Lobad}},
  \bibinfo{author}{\bibfnamefont{R.~D.} \bibnamefont{Averitt}},
  \bibnamefont{and} \bibinfo{author}{\bibfnamefont{A.~J.}
  \bibnamefont{Taylor}}, \bibinfo{journal}{Phys. Rev. B}
  \textbf{\bibinfo{volume}{63}}, \bibinfo{pages}{060410}
  (\bibinfo{year}{2001}).

\bibitem[{\citenamefont{Tobey et~al.}(2008)\citenamefont{Tobey, Prabhakaran,
  Boothroyd, and Cavalleri}}]{Tobey2008PRL}
\bibinfo{author}{\bibfnamefont{R.~I.} \bibnamefont{Tobey}},
  \bibinfo{author}{\bibfnamefont{D.}~\bibnamefont{Prabhakaran}},
  \bibinfo{author}{\bibfnamefont{A.~T.} \bibnamefont{Boothroyd}},
  \bibnamefont{and}
  \bibinfo{author}{\bibfnamefont{A.}~\bibnamefont{Cavalleri}},
  \bibinfo{journal}{Phys. Rev. Lett.} \textbf{\bibinfo{volume}{101}},
  \bibinfo{pages}{197404} (\bibinfo{year}{2008}).

\bibitem[{\citenamefont{Okimoto et~al.}(2007)\citenamefont{Okimoto, Matsuzaki,
  Tomioka, Kezsmarki, Ogasawara, Matsubara, Okamoto, and
  Tokura}}]{Okimoto2007JPSJ}
\bibinfo{author}{\bibfnamefont{Y.}~\bibnamefont{Okimoto}},
  \bibinfo{author}{\bibfnamefont{H.}~\bibnamefont{Matsuzaki}},
  \bibinfo{author}{\bibfnamefont{Y.}~\bibnamefont{Tomioka}},
  \bibinfo{author}{\bibfnamefont{I.}~\bibnamefont{Kezsmarki}},
  \bibinfo{author}{\bibfnamefont{T.}~\bibnamefont{Ogasawara}},
  \bibinfo{author}{\bibfnamefont{M.}~\bibnamefont{Matsubara}},
  \bibinfo{author}{\bibfnamefont{H.}~\bibnamefont{Okamoto}}, \bibnamefont{and}
  \bibinfo{author}{\bibfnamefont{Y.}~\bibnamefont{Tokura}},
  \bibinfo{journal}{J. Phys. Soc. Jap.} \textbf{\bibinfo{volume}{76}},
  \bibinfo{pages}{043702} (\bibinfo{year}{2007}).

\bibitem[{\citenamefont{Prasankumar et~al.}(2007)\citenamefont{Prasankumar,
  Zvyagin, Kamenev, Balakrishnan, Paul, Taylor, and
  Averitt}}]{Prasankumar2007PRB}
\bibinfo{author}{\bibfnamefont{R.~P.} \bibnamefont{Prasankumar}},
  \bibinfo{author}{\bibfnamefont{S.}~\bibnamefont{Zvyagin}},
  \bibinfo{author}{\bibfnamefont{K.~V.} \bibnamefont{Kamenev}},
  \bibinfo{author}{\bibfnamefont{G.}~\bibnamefont{Balakrishnan}},
  \bibinfo{author}{\bibfnamefont{D.~M.} \bibnamefont{Paul}},
  \bibinfo{author}{\bibfnamefont{A.~J.} \bibnamefont{Taylor}},
  \bibnamefont{and} \bibinfo{author}{\bibfnamefont{R.~D.}
  \bibnamefont{Averitt}}, \bibinfo{journal}{Phys. Rev. B}
  \textbf{\bibinfo{volume}{76}}, \bibinfo{pages}{020402}
  (\bibinfo{year}{2007}).

\bibitem[{\citenamefont{Ren et~al.}(2008)\citenamefont{Ren, Ebrahim, Zhao,
  L\"upke, Xu, Adyam, and Li}}]{Ren2008PRB}
\bibinfo{author}{\bibfnamefont{Y.~H.} \bibnamefont{Ren}},
  \bibinfo{author}{\bibfnamefont{M.}~\bibnamefont{Ebrahim}},
  \bibinfo{author}{\bibfnamefont{H.~B.} \bibnamefont{Zhao}},
  \bibinfo{author}{\bibfnamefont{G.}~\bibnamefont{L\"upke}},
  \bibinfo{author}{\bibfnamefont{Z.~A.} \bibnamefont{Xu}},
  \bibinfo{author}{\bibfnamefont{V.}~\bibnamefont{Adyam}}, \bibnamefont{and}
  \bibinfo{author}{\bibfnamefont{Q.}~\bibnamefont{Li}}, \bibinfo{journal}{Phys.
  Rev. B} \textbf{\bibinfo{volume}{78}}, \bibinfo{pages}{014408}
  (\bibinfo{year}{2008}).

\bibitem[{\citenamefont{Li et~al.}(2013)\citenamefont{Li, Patz, Mouchliadis,
  Yan, Lograsso, Perakis, and Wang}}]{Li2013Nature}
\bibinfo{author}{\bibfnamefont{T.}~\bibnamefont{Li}},
  \bibinfo{author}{\bibfnamefont{A.}~\bibnamefont{Patz}},
  \bibinfo{author}{\bibfnamefont{L.}~\bibnamefont{Mouchliadis}},
  \bibinfo{author}{\bibfnamefont{J.}~\bibnamefont{Yan}},
  \bibinfo{author}{\bibfnamefont{T.~A.} \bibnamefont{Lograsso}},
  \bibinfo{author}{\bibfnamefont{I.~E.} \bibnamefont{Perakis}},
  \bibnamefont{and} \bibinfo{author}{\bibfnamefont{J.}~\bibnamefont{Wang}},
  \bibinfo{journal}{Nature} \textbf{\bibinfo{volume}{496}}, \bibinfo{pages}{69}
  (\bibinfo{year}{2013}).

\bibitem[{\citenamefont{Ogasawara et~al.}(2003)\citenamefont{Ogasawara,
  Matsubara, Tomioka, Kuwata-Gonokami, Okamoto, and Tokura}}]{Ogasawar2003PRB}
\bibinfo{author}{\bibfnamefont{T.}~\bibnamefont{Ogasawara}},
  \bibinfo{author}{\bibfnamefont{M.}~\bibnamefont{Matsubara}},
  \bibinfo{author}{\bibfnamefont{Y.}~\bibnamefont{Tomioka}},
  \bibinfo{author}{\bibfnamefont{M.}~\bibnamefont{Kuwata-Gonokami}},
  \bibinfo{author}{\bibfnamefont{H.}~\bibnamefont{Okamoto}}, \bibnamefont{and}
  \bibinfo{author}{\bibfnamefont{Y.}~\bibnamefont{Tokura}},
  \bibinfo{journal}{Phys. Rev. B} \textbf{\bibinfo{volume}{68}},
  \bibinfo{pages}{180407} (\bibinfo{year}{2003}).

\bibitem[{\citenamefont{Millis et~al.}(1996{\natexlab{a}})\citenamefont{Millis,
  Mueller, and Shraiman}}]{Millis1996PRB}
\bibinfo{author}{\bibfnamefont{A.~J.} \bibnamefont{Millis}},
  \bibinfo{author}{\bibfnamefont{R.}~\bibnamefont{Mueller}}, \bibnamefont{and}
  \bibinfo{author}{\bibfnamefont{B.~I.} \bibnamefont{Shraiman}},
  \bibinfo{journal}{Phys. Rev. B} \textbf{\bibinfo{volume}{54}},
  \bibinfo{pages}{5405} (\bibinfo{year}{1996}{\natexlab{a}}).

\bibitem[{\citenamefont{Sheu et~al.}(2011)\citenamefont{Sheu, Trigo, Chien,
  Uher, Arms, Peterson, Walko, Landahl, Chen, Ghimire et~al.}}]{Sheu2011SSC}
\bibinfo{author}{\bibfnamefont{Y.}~\bibnamefont{Sheu}},
  \bibinfo{author}{\bibfnamefont{M.}~\bibnamefont{Trigo}},
  \bibinfo{author}{\bibfnamefont{Y.}~\bibnamefont{Chien}},
  \bibinfo{author}{\bibfnamefont{C.}~\bibnamefont{Uher}},
  \bibinfo{author}{\bibfnamefont{D.}~\bibnamefont{Arms}},
  \bibinfo{author}{\bibfnamefont{E.}~\bibnamefont{Peterson}},
  \bibinfo{author}{\bibfnamefont{D.}~\bibnamefont{Walko}},
  \bibinfo{author}{\bibfnamefont{E.}~\bibnamefont{Landahl}},
  \bibinfo{author}{\bibfnamefont{J.}~\bibnamefont{Chen}},
  \bibinfo{author}{\bibfnamefont{S.}~\bibnamefont{Ghimire}},
  \bibnamefont{et~al.}, \bibinfo{journal}{Solid State Commun.}
  \textbf{\bibinfo{volume}{151}}, \bibinfo{pages}{826} (\bibinfo{year}{2011}).

\bibitem[{\citenamefont{Prasankumar and Taylor}(2011)}]{Rohit2011Book}
\bibinfo{author}{\bibfnamefont{R.~P.} \bibnamefont{Prasankumar}}
  \bibnamefont{and} \bibinfo{author}{\bibfnamefont{A.~J.}
  \bibnamefont{Taylor}}, \emph{\bibinfo{title}{Optical Techniques for
  Solid-State Materials Characterization}} (\bibinfo{publisher}{CRC Press},
  \bibinfo{address}{Boca Raton, Florida}, \bibinfo{year}{2011}),
  chap.~\bibinfo{chapter}{9}.

\bibitem[{\citenamefont{Okimoto et~al.}(1995)\citenamefont{Okimoto, Katsufuji,
  Ishikawa, Urushibara, Arima, and Tokura}}]{Okimoto1995PRL}
\bibinfo{author}{\bibfnamefont{Y.}~\bibnamefont{Okimoto}},
  \bibinfo{author}{\bibfnamefont{T.}~\bibnamefont{Katsufuji}},
  \bibinfo{author}{\bibfnamefont{T.}~\bibnamefont{Ishikawa}},
  \bibinfo{author}{\bibfnamefont{A.}~\bibnamefont{Urushibara}},
  \bibinfo{author}{\bibfnamefont{T.}~\bibnamefont{Arima}}, \bibnamefont{and}
  \bibinfo{author}{\bibfnamefont{Y.}~\bibnamefont{Tokura}},
  \bibinfo{journal}{Phys. Rev. Lett.} \textbf{\bibinfo{volume}{75}},
  \bibinfo{pages}{109} (\bibinfo{year}{1995}).

\bibitem[{\citenamefont{Lu et~al.}(2000)\citenamefont{Lu, Wang, Kwon, and
  Jia}}]{Lu2000Sample}
\bibinfo{author}{\bibfnamefont{C.~J.} \bibnamefont{Lu}},
  \bibinfo{author}{\bibfnamefont{Z.~L.} \bibnamefont{Wang}},
  \bibinfo{author}{\bibfnamefont{C.}~\bibnamefont{Kwon}}, \bibnamefont{and}
  \bibinfo{author}{\bibfnamefont{Q.~X.} \bibnamefont{Jia}},
  \bibinfo{journal}{J. Appl. Phys.} \textbf{\bibinfo{volume}{88}},
  \bibinfo{pages}{4032} (\bibinfo{year}{2000}).

\bibitem[{\citenamefont{Bielecki et~al.}(2010)\citenamefont{Bielecki, Rauer,
  Zanghellini, Gunnarsson, D\"orr, and B\"orjesson}}]{Bielecki2010PRB}
\bibinfo{author}{\bibfnamefont{J.}~\bibnamefont{Bielecki}},
  \bibinfo{author}{\bibfnamefont{R.}~\bibnamefont{Rauer}},
  \bibinfo{author}{\bibfnamefont{E.}~\bibnamefont{Zanghellini}},
  \bibinfo{author}{\bibfnamefont{R.}~\bibnamefont{Gunnarsson}},
  \bibinfo{author}{\bibfnamefont{K.}~\bibnamefont{D\"orr}}, \bibnamefont{and}
  \bibinfo{author}{\bibfnamefont{L.}~\bibnamefont{B\"orjesson}},
  \bibinfo{journal}{Phys. Rev. B} \textbf{\bibinfo{volume}{81}},
  \bibinfo{pages}{064434} (\bibinfo{year}{2010}).

\bibitem[{Thi()}]{ThicknessNote}
\bibinfo{note}{Detailed experimental results and analysis for various film
  thicknesses, shedding light on the origin of the faster thermal decay and
  spin-lattice relaxation in the linear response region, are discussed in
  detail in a separate manuscript in preparation.}

\bibitem[{\citenamefont{Kim et~al.}(1998)\citenamefont{Kim, Jung, and
  Noh}}]{Kim1998PRL}
\bibinfo{author}{\bibfnamefont{K.~H.} \bibnamefont{Kim}},
  \bibinfo{author}{\bibfnamefont{J.~H.} \bibnamefont{Jung}}, \bibnamefont{and}
  \bibinfo{author}{\bibfnamefont{T.~W.} \bibnamefont{Noh}},
  \bibinfo{journal}{Phys. Rev. Lett.} \textbf{\bibinfo{volume}{81}},
  \bibinfo{pages}{1517} (\bibinfo{year}{1998}).

\bibitem[{\citenamefont{Quijada et~al.}(1998)\citenamefont{Quijada,
  \ifmmode~\check{C}\else \v{C}\fi{}erne, Simpson, Drew, Ahn, Millis,
  Shreekala, Ramesh, Rajeswari, and Venkatesan}}]{Quijada1998PRB}
\bibinfo{author}{\bibfnamefont{M.}~\bibnamefont{Quijada}},
  \bibinfo{author}{\bibfnamefont{J.}~\bibnamefont{\ifmmode~\check{C}\else
  \v{C}\fi{}erne}}, \bibinfo{author}{\bibfnamefont{J.~R.}
  \bibnamefont{Simpson}}, \bibinfo{author}{\bibfnamefont{H.~D.}
  \bibnamefont{Drew}}, \bibinfo{author}{\bibfnamefont{K.~H.}
  \bibnamefont{Ahn}}, \bibinfo{author}{\bibfnamefont{A.~J.}
  \bibnamefont{Millis}},
  \bibinfo{author}{\bibfnamefont{R.}~\bibnamefont{Shreekala}},
  \bibinfo{author}{\bibfnamefont{R.}~\bibnamefont{Ramesh}},
  \bibinfo{author}{\bibfnamefont{M.}~\bibnamefont{Rajeswari}},
  \bibnamefont{and}
  \bibinfo{author}{\bibfnamefont{T.}~\bibnamefont{Venkatesan}},
  \bibinfo{journal}{Phys. Rev. B} \textbf{\bibinfo{volume}{58}},
  \bibinfo{pages}{16093} (\bibinfo{year}{1998}).

\bibitem[{\citenamefont{Kaplan et~al.}(1996)\citenamefont{Kaplan, Quijada,
  Drew, Tanner, Xiong, Ramesh, Kwon, and Venkatesan}}]{Kaplan1996PRL}
\bibinfo{author}{\bibfnamefont{S.~G.} \bibnamefont{Kaplan}},
  \bibinfo{author}{\bibfnamefont{M.}~\bibnamefont{Quijada}},
  \bibinfo{author}{\bibfnamefont{H.~D.} \bibnamefont{Drew}},
  \bibinfo{author}{\bibfnamefont{D.~B.} \bibnamefont{Tanner}},
  \bibinfo{author}{\bibfnamefont{G.~C.} \bibnamefont{Xiong}},
  \bibinfo{author}{\bibfnamefont{R.}~\bibnamefont{Ramesh}},
  \bibinfo{author}{\bibfnamefont{C.}~\bibnamefont{Kwon}}, \bibnamefont{and}
  \bibinfo{author}{\bibfnamefont{T.}~\bibnamefont{Venkatesan}},
  \bibinfo{journal}{Phys. Rev. Lett.} \textbf{\bibinfo{volume}{77}},
  \bibinfo{pages}{2081} (\bibinfo{year}{1996}).

\bibitem[{\citenamefont{Hartinger et~al.}(2004)\citenamefont{Hartinger, Mayr,
  Deisenhofer, Loidl, and Kopp}}]{Hartinger2004PRB}
\bibinfo{author}{\bibfnamefont{C.}~\bibnamefont{Hartinger}},
  \bibinfo{author}{\bibfnamefont{F.}~\bibnamefont{Mayr}},
  \bibinfo{author}{\bibfnamefont{J.}~\bibnamefont{Deisenhofer}},
  \bibinfo{author}{\bibfnamefont{A.}~\bibnamefont{Loidl}}, \bibnamefont{and}
  \bibinfo{author}{\bibfnamefont{T.}~\bibnamefont{Kopp}},
  \bibinfo{journal}{Phys. Rev. B} \textbf{\bibinfo{volume}{69}},
  \bibinfo{pages}{100403} (\bibinfo{year}{2004}).

\bibitem[{\citenamefont{Millis et~al.}(1996{\natexlab{b}})\citenamefont{Millis,
  Shraiman, and Mueller}}]{Millis1996PRL}
\bibinfo{author}{\bibfnamefont{A.~J.} \bibnamefont{Millis}},
  \bibinfo{author}{\bibfnamefont{B.~I.} \bibnamefont{Shraiman}},
  \bibnamefont{and} \bibinfo{author}{\bibfnamefont{R.}~\bibnamefont{Mueller}},
  \bibinfo{journal}{Phys. Rev. Lett.} \textbf{\bibinfo{volume}{77}},
  \bibinfo{pages}{175} (\bibinfo{year}{1996}{\natexlab{b}}).

\bibitem[{\citenamefont{Tomioka et~al.}(1996)\citenamefont{Tomioka, Asamitsu,
  Kuwahara, Moritomo, and Tokura}}]{Tomioka1996PRB}
\bibinfo{author}{\bibfnamefont{Y.}~\bibnamefont{Tomioka}},
  \bibinfo{author}{\bibfnamefont{A.}~\bibnamefont{Asamitsu}},
  \bibinfo{author}{\bibfnamefont{H.}~\bibnamefont{Kuwahara}},
  \bibinfo{author}{\bibfnamefont{Y.}~\bibnamefont{Moritomo}}, \bibnamefont{and}
  \bibinfo{author}{\bibfnamefont{Y.}~\bibnamefont{Tokura}},
  \bibinfo{journal}{Phys. Rev. B} \textbf{\bibinfo{volume}{53}},
  \bibinfo{pages}{R1689} (\bibinfo{year}{1996}).

\bibitem[{\citenamefont{Okimoto et~al.}(1998)\citenamefont{Okimoto, Tomioka,
  Onose, Otsuka, and Tokura}}]{Okimoto1998PRB}
\bibinfo{author}{\bibfnamefont{Y.}~\bibnamefont{Okimoto}},
  \bibinfo{author}{\bibfnamefont{Y.}~\bibnamefont{Tomioka}},
  \bibinfo{author}{\bibfnamefont{Y.}~\bibnamefont{Onose}},
  \bibinfo{author}{\bibfnamefont{Y.}~\bibnamefont{Otsuka}}, \bibnamefont{and}
  \bibinfo{author}{\bibfnamefont{Y.}~\bibnamefont{Tokura}},
  \bibinfo{journal}{Phys. Rev. B} \textbf{\bibinfo{volume}{57}},
  \bibinfo{pages}{R9377} (\bibinfo{year}{1998}).

\bibitem[{\citenamefont{Tokura and Nagaosa}(2000)}]{Tokura2000Science}
\bibinfo{author}{\bibfnamefont{Y.}~\bibnamefont{Tokura}} \bibnamefont{and}
  \bibinfo{author}{\bibfnamefont{N.}~\bibnamefont{Nagaosa}},
  \bibinfo{journal}{Science} \textbf{\bibinfo{volume}{288}},
  \bibinfo{pages}{462} (\bibinfo{year}{2000}).

\bibitem[{\citenamefont{Jung et~al.}(2000)\citenamefont{Jung, Lee, Noh, Choi,
  Moritomo, Wang, and Wei}}]{Jung2000PRB}
\bibinfo{author}{\bibfnamefont{J.~H.} \bibnamefont{Jung}},
  \bibinfo{author}{\bibfnamefont{H.~J.} \bibnamefont{Lee}},
  \bibinfo{author}{\bibfnamefont{T.~W.} \bibnamefont{Noh}},
  \bibinfo{author}{\bibfnamefont{E.~J.} \bibnamefont{Choi}},
  \bibinfo{author}{\bibfnamefont{Y.}~\bibnamefont{Moritomo}},
  \bibinfo{author}{\bibfnamefont{Y.~J.} \bibnamefont{Wang}}, \bibnamefont{and}
  \bibinfo{author}{\bibfnamefont{X.}~\bibnamefont{Wei}},
  \bibinfo{journal}{Phys. Rev. B} \textbf{\bibinfo{volume}{62}},
  \bibinfo{pages}{481} (\bibinfo{year}{2000}).

\bibitem[{\citenamefont{Millis}(1998)}]{Millis1998Nature}
\bibinfo{author}{\bibfnamefont{A.~J.} \bibnamefont{Millis}},
  \bibinfo{journal}{Science} \textbf{\bibinfo{volume}{392}},
  \bibinfo{pages}{147} (\bibinfo{year}{1998}).

\bibitem[{\citenamefont{Averitt and Taylor}(2002)}]{Averitt2002JPCM}
\bibinfo{author}{\bibfnamefont{R.~D.} \bibnamefont{Averitt}} \bibnamefont{and}
  \bibinfo{author}{\bibfnamefont{A.~J.} \bibnamefont{Taylor}},
  \bibinfo{journal}{J. Phys.: Condens. Matter} \textbf{\bibinfo{volume}{14}},
  \bibinfo{pages}{R1357} (\bibinfo{year}{2002}).

\bibitem[{\citenamefont{Sheu et~al.}(2012)\citenamefont{Sheu, Trugman, Park,
  Lee, Yi, Cheong, Jia, Taylor, and Prasankumar}}]{Sheu2012APL}
\bibinfo{author}{\bibfnamefont{Y.~M.} \bibnamefont{Sheu}},
  \bibinfo{author}{\bibfnamefont{S.~A.} \bibnamefont{Trugman}},
  \bibinfo{author}{\bibfnamefont{Y.-S.} \bibnamefont{Park}},
  \bibinfo{author}{\bibfnamefont{S.}~\bibnamefont{Lee}},
  \bibinfo{author}{\bibfnamefont{H.~T.} \bibnamefont{Yi}},
  \bibinfo{author}{\bibfnamefont{S.-W.} \bibnamefont{Cheong}},
  \bibinfo{author}{\bibfnamefont{Q.~X.} \bibnamefont{Jia}},
  \bibinfo{author}{\bibfnamefont{A.~J.} \bibnamefont{Taylor}},
  \bibnamefont{and} \bibinfo{author}{\bibfnamefont{R.~P.}
  \bibnamefont{Prasankumar}}, \bibinfo{journal}{Appl. Phys. Lett.}
  \textbf{\bibinfo{volume}{100}}, \bibinfo{pages}{242904}
  (\bibinfo{year}{2012}).

\bibitem[{\citenamefont{Wen et~al.}(2013)\citenamefont{Wen, Chen, Cosgriff,
  Walko, Lee, Adamo, Schaller, Ihlefeld, Dufresne, Schlom et~al.}}]{Wen2013PRL}
\bibinfo{author}{\bibfnamefont{H.}~\bibnamefont{Wen}},
  \bibinfo{author}{\bibfnamefont{P.}~\bibnamefont{Chen}},
  \bibinfo{author}{\bibfnamefont{M.~P.} \bibnamefont{Cosgriff}},
  \bibinfo{author}{\bibfnamefont{D.~A.} \bibnamefont{Walko}},
  \bibinfo{author}{\bibfnamefont{J.~H.} \bibnamefont{Lee}},
  \bibinfo{author}{\bibfnamefont{C.}~\bibnamefont{Adamo}},
  \bibinfo{author}{\bibfnamefont{R.~D.} \bibnamefont{Schaller}},
  \bibinfo{author}{\bibfnamefont{J.~F.} \bibnamefont{Ihlefeld}},
  \bibinfo{author}{\bibfnamefont{E.~M.} \bibnamefont{Dufresne}},
  \bibinfo{author}{\bibfnamefont{D.~G.} \bibnamefont{Schlom}},
  \bibnamefont{et~al.}, \bibinfo{journal}{Phys. Rev. Lett.}
  \textbf{\bibinfo{volume}{110}}, \bibinfo{pages}{037601}
  (\bibinfo{year}{2013}).

\bibitem[{mul()}]{multiRNote}
\bibinfo{note}{We note that potential multiple reflections in these
  heterostructures will only affect the magnitude of the time-integrated DC
  reflectivity, without influencing the time-resolved signal, particularly
  since only the properties of LCMO are directly changing with time after
  photoexcitation.}

\bibitem[{\citenamefont{Kundys et~al.}(2010)\citenamefont{Kundys, Viret,
  Colson, and Kundys}}]{Kundys2010NM}
\bibinfo{author}{\bibfnamefont{B.}~\bibnamefont{Kundys}},
  \bibinfo{author}{\bibfnamefont{M.}~\bibnamefont{Viret}},
  \bibinfo{author}{\bibfnamefont{D.}~\bibnamefont{Colson}}, \bibnamefont{and}
  \bibinfo{author}{\bibfnamefont{D.~O.} \bibnamefont{Kundys}},
  \bibinfo{journal}{Nat. Mater.} \textbf{\bibinfo{volume}{9}},
  \bibinfo{pages}{803} (\bibinfo{year}{2010}).

\bibitem[{Spi()}]{SpinNote}
\bibinfo{note}{In G-type BFO, spins are aligned perpendicular to the FE
  polarization (111) and the equivalent variants. The spin alignment can be
  projected into in-plane and out-of-plane components. The induced AFM order in
  LCMO, however, depends mainly on the out-of-plane spin components of BFO
  since only the perpendicular magnetic susceptibility ($\chi_{\perp}$)
  dominates the AFM order, i.e. only a perpendicular field can cant the
  N\'{e}el axis.}

\bibitem[{Sin()}]{Singh2013unpublished}
\bibinfo{note}{Singh, S. et al., unpublished work.}

\bibitem[{Swi()}]{SwitchNote}
\bibinfo{note}{In the Goodenough-Kanamori-Anderson picture, the magnetic
  interaction is proportional to the overlap squared of the Wannier
  wavefunctions of the two atoms
  \cite{Anderson1959PR,Goodenough1960PR,Kanamori1963PTP}. In an insulator or
  the vacuum, the overlap decays exponentially \cite{Wannier1937PR}. In a
  metal, from arguments similar to RKKY
  \cite{Ruderma1954PR,Kasuya1956PTP,Yosida1957PR}, a power law is more typical.
  In this case, $l=l_1+l_2$, where $l$ is the separation between Fe and Mn
  ions, $l_1$ the part on the metallic side, and $l_2$ the part on the
  insulating side. If one wants to know how the magnetic interaction $J$
  depends on $l$, one could ask whether $l$ changes mainly because $l_1$
  changes, or $l_2$, or both. In either case, $J$ will decrease significantly
  with increasing $l$. If it is mainly due to changes in $l_2$, as expected
  here, one expects an exponential dependence.}

\bibitem[{\citenamefont{Chang et~al.}(2011)\citenamefont{Chang, Kalinin,
  Morozovska, Huijben, Chu, Yu, Ramesh, Eliseev, Svechnikov, Pennycook
  et~al.}}]{Chang2011AM}
\bibinfo{author}{\bibfnamefont{H.~J.} \bibnamefont{Chang}},
  \bibinfo{author}{\bibfnamefont{S.~V.} \bibnamefont{Kalinin}},
  \bibinfo{author}{\bibfnamefont{A.~N.} \bibnamefont{Morozovska}},
  \bibinfo{author}{\bibfnamefont{M.}~\bibnamefont{Huijben}},
  \bibinfo{author}{\bibfnamefont{Y.-H.} \bibnamefont{Chu}},
  \bibinfo{author}{\bibfnamefont{P.}~\bibnamefont{Yu}},
  \bibinfo{author}{\bibfnamefont{R.}~\bibnamefont{Ramesh}},
  \bibinfo{author}{\bibfnamefont{E.~A.} \bibnamefont{Eliseev}},
  \bibinfo{author}{\bibfnamefont{G.~S.} \bibnamefont{Svechnikov}},
  \bibinfo{author}{\bibfnamefont{S.~J.} \bibnamefont{Pennycook}},
  \bibnamefont{et~al.}, \bibinfo{journal}{Adv. Mater.}
  \textbf{\bibinfo{volume}{23}}, \bibinfo{pages}{2474} (\bibinfo{year}{2011}).

\bibitem[{\citenamefont{Basov et~al.}(2011)\citenamefont{Basov, Averitt,
  van~der Marel, Dressel, and Haule}}]{BasovReview2011}
\bibinfo{author}{\bibfnamefont{D.~N.} \bibnamefont{Basov}},
  \bibinfo{author}{\bibfnamefont{R.~D.} \bibnamefont{Averitt}},
  \bibinfo{author}{\bibfnamefont{D.}~\bibnamefont{van~der Marel}},
  \bibinfo{author}{\bibfnamefont{M.}~\bibnamefont{Dressel}}, \bibnamefont{and}
  \bibinfo{author}{\bibfnamefont{K.}~\bibnamefont{Haule}},
  \bibinfo{journal}{Rev. Mod. Phys.} \textbf{\bibinfo{volume}{83}},
  \bibinfo{pages}{471} (\bibinfo{year}{2011}).

\bibitem[{\citenamefont{Hopkins et~al.}(2013)\citenamefont{Hopkins, Adamo, Ye,
  Huey, Lee, Schlom, and Ihlefeld}}]{Hopkins2013APL}
\bibinfo{author}{\bibfnamefont{P.~E.} \bibnamefont{Hopkins}},
  \bibinfo{author}{\bibfnamefont{C.}~\bibnamefont{Adamo}},
  \bibinfo{author}{\bibfnamefont{L.}~\bibnamefont{Ye}},
  \bibinfo{author}{\bibfnamefont{B.~D.} \bibnamefont{Huey}},
  \bibinfo{author}{\bibfnamefont{S.~R.} \bibnamefont{Lee}},
  \bibinfo{author}{\bibfnamefont{D.~G.} \bibnamefont{Schlom}},
  \bibnamefont{and} \bibinfo{author}{\bibfnamefont{J.~F.}
  \bibnamefont{Ihlefeld}}, \bibinfo{journal}{Appl. Phys. Lett.}
  \textbf{\bibinfo{volume}{102}}, \bibinfo{eid}{121903} (\bibinfo{year}{2013}).

\bibitem[{\citenamefont{Sheu et~al.}(2013)\citenamefont{Sheu, Trugman, Xiong,
  Park, Lee, Yi, Cheong, Jia, Taylor, and Prasankumar}}]{Sheu2013EPJ}
\bibinfo{author}{\bibfnamefont{Y.~M.} \bibnamefont{Sheu}},
  \bibinfo{author}{\bibfnamefont{S.~A.} \bibnamefont{Trugman}},
  \bibinfo{author}{\bibfnamefont{J.}~\bibnamefont{Xiong}},
  \bibinfo{author}{\bibfnamefont{Y.-S.} \bibnamefont{Park}},
  \bibinfo{author}{\bibfnamefont{S.}~\bibnamefont{Lee}},
  \bibinfo{author}{\bibfnamefont{H.~T.} \bibnamefont{Yi}},
  \bibinfo{author}{\bibfnamefont{S.-W.} \bibnamefont{Cheong}},
  \bibinfo{author}{\bibfnamefont{Q.~X.} \bibnamefont{Jia}},
  \bibinfo{author}{\bibfnamefont{A.~J.} \bibnamefont{Taylor}},
  \bibnamefont{and} \bibinfo{author}{\bibfnamefont{R.~P.}
  \bibnamefont{Prasankumar}}, \bibinfo{journal}{EPJ Web of Conferences}
  \textbf{\bibinfo{volume}{41}}, \bibinfo{pages}{03018} (\bibinfo{year}{2013}).

\bibitem[{\citenamefont{Anderson}(1959)}]{Anderson1959PR}
\bibinfo{author}{\bibfnamefont{P.~W.} \bibnamefont{Anderson}},
  \bibinfo{journal}{Phys. Rev.} \textbf{\bibinfo{volume}{115}},
  \bibinfo{pages}{2} (\bibinfo{year}{1959}).

\bibitem[{\citenamefont{Goodenough}(1960)}]{Goodenough1960PR}
\bibinfo{author}{\bibfnamefont{J.~B.} \bibnamefont{Goodenough}},
  \bibinfo{journal}{Phys. Rev.} \textbf{\bibinfo{volume}{120}},
  \bibinfo{pages}{67} (\bibinfo{year}{1960}).

\bibitem[{\citenamefont{Kanamori}(1963)}]{Kanamori1963PTP}
\bibinfo{author}{\bibfnamefont{J.}~\bibnamefont{Kanamori}},
  \bibinfo{journal}{Prog. Theor. Phys.} \textbf{\bibinfo{volume}{30}},
  \bibinfo{pages}{275} (\bibinfo{year}{1963}).

\bibitem[{\citenamefont{Wannier}(1937)}]{Wannier1937PR}
\bibinfo{author}{\bibfnamefont{G.~H.} \bibnamefont{Wannier}},
  \bibinfo{journal}{Phys. Rev.} \textbf{\bibinfo{volume}{52}},
  \bibinfo{pages}{191} (\bibinfo{year}{1937}).

\bibitem[{\citenamefont{Ruderman and Kittel}(1954)}]{Ruderma1954PR}
\bibinfo{author}{\bibfnamefont{M.~A.} \bibnamefont{Ruderman}} \bibnamefont{and}
  \bibinfo{author}{\bibfnamefont{C.}~\bibnamefont{Kittel}},
  \bibinfo{journal}{Phys. Rev.} \textbf{\bibinfo{volume}{96}},
  \bibinfo{pages}{99} (\bibinfo{year}{1954}).

\bibitem[{\citenamefont{Kasuya}(1956)}]{Kasuya1956PTP}
\bibinfo{author}{\bibfnamefont{T.}~\bibnamefont{Kasuya}},
  \bibinfo{journal}{Prog. Theor. Phys.} \textbf{\bibinfo{volume}{16}}
  (\bibinfo{year}{1956}).

\bibitem[{\citenamefont{Yosida}(1957)}]{Yosida1957PR}
\bibinfo{author}{\bibfnamefont{K.}~\bibnamefont{Yosida}},
  \bibinfo{journal}{Phys. Rev.} \textbf{\bibinfo{volume}{106}},
  \bibinfo{pages}{893} (\bibinfo{year}{1957}).

\end{thebibliography}

\end{document}